%% file: ClimRecon.tex
\documentclass[letterpaper,12pt]{article}

\usepackage{graphicx} 
\usepackage{amsmath}
\usepackage{amsthm}
\usepackage{amsfonts}
\usepackage{bm}
\usepackage{amssymb}
\usepackage{lineno}
\usepackage{siunitx}
\usepackage{xspace}
\usepackage[]{natbib}
\bibpunct{(}{)}{,}{a}{}{,}
\usepackage{url}
\usepackage{tabularx}

\usepackage{epic} 
\usepackage{ecltree} 
\usepackage{pstricks} 
\usepackage{pst-node} 
\usepackage{psfrag} 

\usepackage{caption}
\usepackage{subcaption}

\usepackage{setspace}

\newcommand{\tsinv}{\textsc{ts\_inv}\xspace}
\newcommand{\rcsinv}{\textsc{rcs\_inv}\xspace}
\newcommand{\tsclass}{\textsc{ts\_class}\xspace}
\newcommand{\rcsclass}{\textsc{rcs\_class}\xspace}
\newcommand{\mtscon}{\textsc{mb\_ts\_con}\xspace}
\newcommand{\mrcscon}{\textsc{mb\_rcs\_con}\xspace}

\newcommand{\mtsspl}{\textsc{mb\_ts\_spl}\xspace}
\newcommand{\mtssplhard}{\textsc{mb\_ts\_spl\_pl}\xspace}
\newcommand{\mtssplsoft}{\textsc{mb\_ts\_spl\_ip}\xspace}

\begin{document}

\allowdisplaybreaks

\title{A Model-Based Approach to Climate Reconstruction Using Tree-Ring Data\thanks{To appear in the \textit{Journal of the American Statistical Association.}}}
\author{Matthew R. Schofield$^{12}$\thanks{E-mail: \texttt{mschofield@maths.otago.ac.nz}}, Richard J. Barker$^{2}$, Andrew Gelman$^{3}$,\\ Edward R. Cook$^{4}$ and Keith R. Briffa$^{5}$\\
  \normalsize{$^{1}$Department of Statistics, University of Kentucky, Lexington, KY, USA.}\\
  \normalsize{$^{2}$Department of Mathematics and Statistics, University of Otago,}\\
  \normalsize{PO Box 56, Dunedin, New Zealand.}\\
  \normalsize{$^{3}$Department of Statistics, Columbia University, NY, USA.}\\
  \normalsize{$^{4}$Lamont--Doherty Earth Observatory, Palisades, NY, USA.}\\
  \normalsize{$^{5}$Climatic Research Unit, School of Environmental Sciences,}\\ 
  \normalsize{University of East Anglia, Norwich NR4 7TJ, UK.}\\
} 
\maketitle

\input{Intro}
\input{DataMeth}
\input{Calib}
\input{models}

\input{modcheck}
\input{discuss}

\input{acknow}



\bibliographystyle{asa} 
\bibliography{biblio}


\include{figures}

\end{document}

%% file: Intro.tex
\begin{abstract}
\input{abstract_mrs}
\end{abstract}

\section{Introduction}\label{sec:intro}
Instrumental meteorological records from the past 200 years support a scientific consensus that climate is changing 
\citep{Trenberth2007}.    What is less clear is how to interpret these changes in an historical context spanning many hundreds, or thousands of years.  The statistical problem is to infer historical climate without having direct observations beyond the most recent two centuries.  A common solution is to index historical climate values using proxy observations from natural archives such as tree rings, lake sediments, bore holes, corals and ice cores (see \cite{Jones2004} and \cite{Jansen2007} for examples).  
Reconstructions based on these sources feature prominently in the Fifth Assessment Report of Working Group 1 of the Intergovernmental Panel on Climate Change \cite[]{Masson-Delmotte2013}.  This report and other work, including that of the Past Global Changes (PAGES) 2K group \cite[]{PAGES2kConsortium2013}, have demonstrated robustness in the general pattern of variability in reconstructions representing large regions at decadal to centennial timescales.

The standard approach to reconstructing climate from tree-ring data involves (i) filtering out any signal from non-climatic sources so that only climate signal (and random noise) remain, (ii) combining multiple filtered tree series from the same macro-environmental setting into one mean chronology,  and (iii) using the resulting chronology, alone or combined with other proxy measurements, to reconstruct local, hemispheric or global climate.  
The first two steps are referred to as ``standardization'' in tree-ring studies \cite[e.g.][]{Fritts1976}.
There is a rich literature describing dendroclimatic methods devoted to preserving the climate signal during  standardization \cite[e.g.][]{Briffa1992,Cook1995,Melvin2008,Esper2009,Briffa2011}. 
However, the major focus of recent research has been on the third step 
\cite[for example, see][]{Jones2009,Tingley2012,Wahl2012}.  This has resulted in several methodological developments for reconstructing climate values from filtered proxy data.  
More recently, Bayesian inference has been proposed.  \cite{Haslett2006} used a Bayesian approach for reconstruction using a pollen proxy. \cite{Li2010}, \cite{Tingley2010,Tingley2010a} and \cite{McShane2011} explore Bayesian approaches for multi-proxy reconstructions.  

The standardization process results in a chronology that is easily shared among the community for use in climate reconstructions.   However, in subsequent use the chronologies are treated as data and the standardization process does not feature in the climate reconstruction model.  One consequence is that uncertainties from steps (1) and (2) above do not propagate through to the final predictions.  Any incorrectly modeled variation from the standardization of the proxy measurements will be carried forward to the climate reconstruction.   The multi-step procedures also make it different to check overall model fit. Assumptions made in the second or third stages cannot be checked against the raw data.
To overcome these problems we merge the three steps above, specifying a joint model for the raw data and climate measure.  Simultaneously describing climatic and non-climatic influences on the raw data makes it possible to overcome difficulties encountered in multi-step procedures, such as reconstructing long-scale climate variability, often referred to as the segment-length curse \cite[]{Cook1995}.

Use of a joint model ensures that assumptions are made explicit.  
As a first step we need to find model-based descriptions for the underlying assumptions in both the standardization and reconstruction processes described above.  Statistical calibration is often used to reconstruct the climate in step (3) above, with different approaches typically presented as a choice between various statistical procedures.  In our framework, the different approaches can be viewed as differences in modeling assumptions and comparison between procedures becomes a choice between models.

Once a modeling framework has been specified, we can consider alternative assumptions and examine how the predictions of historical climate depend on changes in both the standardization and reconstruction components of the model.  
If the predictions change substantially, this shows sensitivity of the modeling assumptions and can help reveal limitations of the data.

We first introduce the data in section \ref{sect:data} and summarize currently used methodology for standardizing tree-ring data in section \ref{sect:stand}.  We then comment on the problems with the current methodology in section \ref{sect:prob} before defining and exploring the model-based approach in section \ref{sect:mod1} and considering model adequacy in section \ref{sect:modcheck}.  We then discuss the implications of the modeling approach in section \ref{sect:discuss}, that includes a summary for the paleoclimate community.

%% file: abstract_mrs.tex
Quantifying long-term historical climate  
is fundamental to understanding recent climate change.  Most instrumentally recorded climate data are only available for the past
200 years, so proxy observations from natural archives are often considered. 
We describe a model-based approach to reconstructing climate defined in terms of raw tree-ring measurement data that simultaneously accounts for non-climatic and climatic variability. 

In this approach we specify a joint model for the tree-ring data and climate variable that we fit using Bayesian inference.  We consider a range of prior densities and   
compare the modeling approach to current methodology using an example case of Scots pine from Tornetr\"{a}sk, Sweden to reconstruct growing season temperature.  We describe how current approaches translate into particular model assumptions.  We explore how changes to various components in the model-based approach affect the resulting reconstruction.

We show that minor changes in model specification can have little effect on model fit but lead to large changes in the predictions.  In particular, the periods of relatively warmer and cooler temperatures are robust between models, but the magnitude of the resulting temperatures are highly model dependent.  
Such sensitivity may not be apparent with traditional approaches because the
underlying statistical model is often hidden or poorly described.  

%% file: DataMeth.tex
\section{Data}\label{sect:data}
Our tree-ring data are records of annual radial growth increments (i.e. ring widths) of living and well-preserved remnant (subfossil) Scots pine (\textit{Pinus sylvestris}) growing near the latitudinal tree-line in Tornetr\"{a}sk, northern Sweden \cite[]{Grudd2002,Briffa2008}.  Ring widths are measured along one or more radii, from the earliest ring at or near the center of the tree bole outward to include the most recently formed growth ring.
They represent the annual radial expansion of woody tissue in the bole of a tree. They are typically measured from samples taken at breast height. While ring widths vary from tree to tree and also at different heights in the same tree, the overall pattern of year to year relative ring width changes is similar.  The absolute dimensions of these ring widths vary according to what part of the tree they represent and to the overall growth rate of the tree in question.  Many trees of the same species are sampled from a site, defined as an area within the same macro-environment, to ensure within site replication.

It is assumed that ring width variation reflects, among other things, climatic variation \cite[]{Fritts1976}.  For the Tornetr\"{a}sk trees it is believed that the important climatic influence is the growing season temperature \cite[]{Grudd2002}.  However, even in an unchanging environment 
 ring widths change over time due to the changing allocation of growth products around the outside of an expanding bole as the tree grows. 
 Trees exhibit various and sometimes complex growth forms but a pattern of early ring width increase followed by a slower reduction in ring widths in older age is generally observed. Because such changes in measured ring width time series do not represent climate variability it is important to account for them in the analysis.  Traditionally, such variations have been filtered out via the standardization process referred to in section \ref{sec:intro}.

It is important that each ring width is assigned to its exact calendar year of formation.  This is done by a process called ``crossdating'' \cite[]{Fritts1976} that matches high-frequency patterns of wide and narrow rings between the sampled trees. This helps identify and enable the correction of dating errors that might arise by simple ring counting \cite[]{Fritts1989}.  More details about crossdating are in the supplementary materials, section 1 (herein S1).  The data we use here are obtained after crossdating.  We discuss potential uncertainty resulting from the crossdating procedure in section \ref{sect:discuss}.
 
The full dataset consists of ring widths from $587$ trees.  We analyze a subset comprising $k=247$ trees with at least $20$ ring width observations after the year $1496$, leading to a $500$ year series.  The number of ring width observations for each tree (defined as the segment length) varied from $26$ to $485$ years, with an average segment length of $179$ years.  Additional details about the data, including plots, are available in S1.  

As the climatic factor limiting growth of these trees is believed to be growing season temperature, the climate variable we reconstruct is the mean temperature during June-August for the Abisko weather station, the closest weather station to Tornetr\"{a}sk.  
Temperature has been directly measured from 1913 to 1995. In all models that follow, we assume that these temperature values are observed without error.  Our goal is to predict the mean June-August temperature values from $1496$ to $1912$.  There are $121$ trees that have 
measurements during the period of observed temperatures.  To account for the non-climatic ring growth, the only other covariate we consider is the age of the tree in each year.  By convention, the age of a tree is defined as the number of years since the innermost ring of the core sample; this corresponds to the year in which the tree reached the height at which the sample was taken for living trees or to the innermost year for the remnant wood samples. 

We use the Tornetr\"{a}sk data for two reasons.  Firstly, the correlation between the growth increments and climate is strong enough to make viable local reconstructions (see section \ref{sect:prob} for details).  Secondly, the spread of age classes of the trees is relatively even through time.  This makes the use of regional curve standardization (discussed in section \ref{sect:RCS}) viable for these data as well as traditional standardization (discussed in section \ref{sect:typical}).

Although our focus here is on reconstructing growing season temperature, the models we fit and the ideas we discuss can be generalized to other climate variables and to global and spatial field based reconstructions.

\subsection{Notation}
The raw data are denoted by the $k \times n$ matrix $\bm{y}$, where $y_{it}$ is the observed ring width for tree $i$ in year $t$.  We use $f_{i}$ and $l_{i}$ to denote the first and last year tree $i$ was measured.  We split $\bm{y}$ into a $k \times m$ matrix $\bm y^{mis}$ and a $k \times (n-m)$ matrix $\bm y^{obs}$.  The former denotes values of $\bm{y}$ for which the climate values $\bm x^{mis} = (x_{1},\ldots,x_{m})$ are missing; the latter denotes values of $\bm{y}$ for which climate values $\bm x^{obs} = (x_{m+1},\ldots,x_{n})$ have been observed.   We denote tree age by the $k \times n$ matrix $\bm a$, where $a_{it}$ is the age of tree $i$ in year $t$.  

Throughout we distinguish between climate and temperature.  When describing general methodology, we refer to $\bm x$ as a climate variable.  When discussing the Tornetr\"{a}sk data, we refer to $\bm x$ as temperature.

\section{Standard methodology}\label{sect:stand}
In this section, we outline the traditional approach to the multi-step procedure for reconstructing climate variables.  
The first step involves standardizing the ring widths, $\bm y$ \cite[]{Fritts1976,Cook1990}.  Standardization serves to (1) filter out non-climatic growth influences, in particular, age effects and tree to tree differences in growth, and (2) compress the information contained in $k$ noisy tree-ring series at a particular site into one synthetic chronology that contains a common signal, assumed to be climatic in nature.
We denote the standardized chronologies obtained from $\bm y$ by the vector $\bm z = (z_{1},\ldots,z_{n})$, where $z_{t}$ is the value for year $t$.  As for $\bm y$, we split $\bm z$ into $\bm z^{mis} = (z_{1},\ldots,z_{m})$ and $\bm z^{obs} = (z_{m+1},\ldots,z_{n})$.  

We now outline two standardization procedures, which we refer to as traditional standardization (TS) and regional curve standardization (RCS).  

\subsection{Traditional standardization}\label{sect:typical}
TS involves selecting and fitting a statistical model to the annual growth increments of each tree separately in order to remove the non-climatic growth influences such as those associated with tree age \cite[]{Fritts1976}.  Raw measurements, $y_{it}$ are then scaled to give an index for each tree, $w_{it} = y_{it}/\hat{y}_{it}$, where $\hat{y}_{it}$ is the fitted value under the assumed model.
This index is then averaged across trees at a site to give a mean index (or chronology) for that site, $z_{t} = m(w_{1t},\ldots,w_{kt})$.  The 
function $m(\cdot)$ is a measure of central location and may be chosen to minimize the effect of outliers.
Variation in $z_{t}$ is presumed to reflect primarily climatic influences with tree-specific growth-related effects removed by scaling.  
Deviations for each tree may be further modeled to remove autocorrelation before being combined \cite[]{Cook1990}.  Almost all datasets in the International Tree-Ring Data Bank (\url{http://www.ncdc.noaa.gov/paleo/treering.html}) have been standardized in this way.  

TS has two major problems.  First, the uncertainty about parameters in the standardization is ignored in all further modeling.  Second, incorrectly modeled variation in the standardization output can distort modeling of the climate signal, inducing bias in the reconstructed climate values.  The problem is that TS effectively detrends each tree-ring series in an attempt to remove the non-climatic growth.  This can easily lead to the unintentional removal of climatic influences on growth from the tree-ring chronology \cite[]{Cook1995}; i.e. some of the low-frequency climate ``baby'' can get thrown out with the non-climate ``bathwater.''  

This second problem has been named the ``segment length curse'' \cite[]{Cook1995}:  one can only expect to recover climate signals that have a higher frequency (in relation to climate cycles) than the inverse of segment length of individual tree-ring measurements.  As low-frequency climate signal (and possibly some medium-frequency signal) is removed during standardization, the examination of changes in climate over hundreds, or thousands of years, becomes problematic.  We use simulated temperature and tree-ring series to demonstrate the problem in S3. 

\subsection{Regional curve standardization}\label{sect:RCS}
RCS represents an attempt to overcome the segment length curse and preserve the low-frequency climate signal \cite[]{Briffa1992}.  
RCS assumes that there is a common growth rate for all trees of the same age (or age class, e.g. we could assume a common growth rate applies to binned ages of 10-20 year increments).  We define $\hat{G}_{j}$ to be the estimated growth rate for age $j$ and obtain an index $w_{it} = y_{it}/\hat{G}_{a_{it}}$.
%
As before, we then average the index to obtain the chronology $z_{t} = m(w_{1t},\ldots,w_{kt})$.


The idea behind RCS is that climatic effects on the standardization curve can be largely eliminated by averaging measurements across many trees sampled from across a range of overlapping time periods; in effect, removing the climate influence on $\hat{G}_{j}$.  The variability in the index is assumed to reflect changes in climatic conditions that can exceed the lengths of the individual tree ring series being detrended, see \cite{Briffa1992} for details.  RCS makes minimal assumptions about the shape of the non-climatic growth trend, although such constraints may be included.  

RCS has several potential problems \cite[]{Briffa2011}.  As with TS, incorrectly modeled variation can distort modeling of the climate signal and induce bias in the resulting reconstruction, e.g. tree-specific differences in the growth can be misinterpreted as changes in the climate signal, particularly when the sample size is small.  Furthermore, the RCS approach also ignores uncertainty about the standardization procedure in all further modeling.

RCS is inappropriate when samples are taken from trees that are 
all about the same age since they form 
a common age class and have been exposed to the same climate conditions.  
Fortunately, many chronologies have a range of ages at each time period, especially when they consist of living and sub-fossil trees.  This is the case for the Tornetr\"{a}sk Scots pine data set which has a relatively even distribution of tree age classes spread over a relatively long time.  This makes simple implementation of the RCS technique a viable option, provided the mean biological age trend can be assumed to be approximately constant between trees.  More complex implementations of RCS have been developed, such as processing subsets of the full data set to allow trees to have different growth rates \cite[]{Melvin2013}.  The analysis we describe here uses a single standardization curve RCS implementation.  
%


%% file: Calib.tex
\subsection{Statistical calibration} \label{sec:StatisticalCalibration}
Following standardization, statistical calibration \citep{Osborne1991} is used to reconstruct climate from the site-specific chronologies.  
We initially focus on univariate linear calibration, where we assume that a single chronology $\bm z$ has a linear relationship with climate observations $\bm x$.  

Observed data $\bm z^{obs}$ and $\bm x^{obs}$ are used to estimate parameters describing the relationship, which are then used to predict $\bm x^{mis}$ using $\bm z^{mis}$.  This can be done using either: (i) classical calibration, where we regress $\bm z^{obs}$ on $\bm x^{obs}$, or (ii) inverse calibration, where $\bm x^{obs}$ is regressed on $\bm z^{obs}$.  The two corresponding procedures, along with their properties are described in S2.  In S2 we show how the underlying model for inverse calibration can be obtained  if (i) we start with the model for classical calibration, (ii) consider $x_{1},\ldots,x_{n}$ to be $iid$ realizations of $n$ normal random variables, and (iii) use Bayes rule to find the  distribution of $x_t$ conditional on $z_t$.
The implied normal model for $\bm x$ leads to predictions\footnote{Throughout we use prediction to refer to making inference regarding $x_{t},~t=1,\ldots,m$.} from  inverse calibration being shrunk toward the mean value of $\bm x^{obs}$ compared with the predictions from classical calibration.  This helps explain why inverse calibration is preferred in terms of MSE only if $x^{mis}_t$ is close to the mean $\bm x^{obs}$ value.   

Extensions to the multivariate case are also discussed in S2.  In particular, we show how the model underlying the RegEM algorithm of \cite{Schneider2001}, commonly used for multiple proxy climate reconstructions, is equivalent to a multivariate extension of inverse calibration.  See \cite{Christiansen2011}, \cite{Christiansen2011a} and \cite{Christiansen2014} for discussion of calibration assumptions with respect to various climate reconstructions.

%% file: models.tex
\section{Problems with existing approaches}\label{sect:prob}
%
We begin by examining the local reconstruction for the Scots pine dataset using traditional methods.  We   
construct TS and RCS chronologies using the approaches outlined in sections \ref{sect:typical} and \ref{sect:RCS}.  We use a negative exponential growth curve in TS to account for tree age and we assume the same growth rate for all trees in binned (10-year) age groups in RCS , with more details provided in S4.  Of interest is how resulting predictions compare with the assumptions and properties of both the standardization approaches, described in section \ref{sect:stand}, as well as the calibration approaches, described in section \ref{sec:StatisticalCalibration} and S2.  We refer to the reconstructions found using the traditional multi-step approach as \tsinv, \tsclass, \rcsinv and \rcsclass.  Table \ref{tab:models} describes the labels for
  all reconstructions we consider.

\begin{figure} [!htbp]
  \centering
   \includegraphics[width =\textwidth]{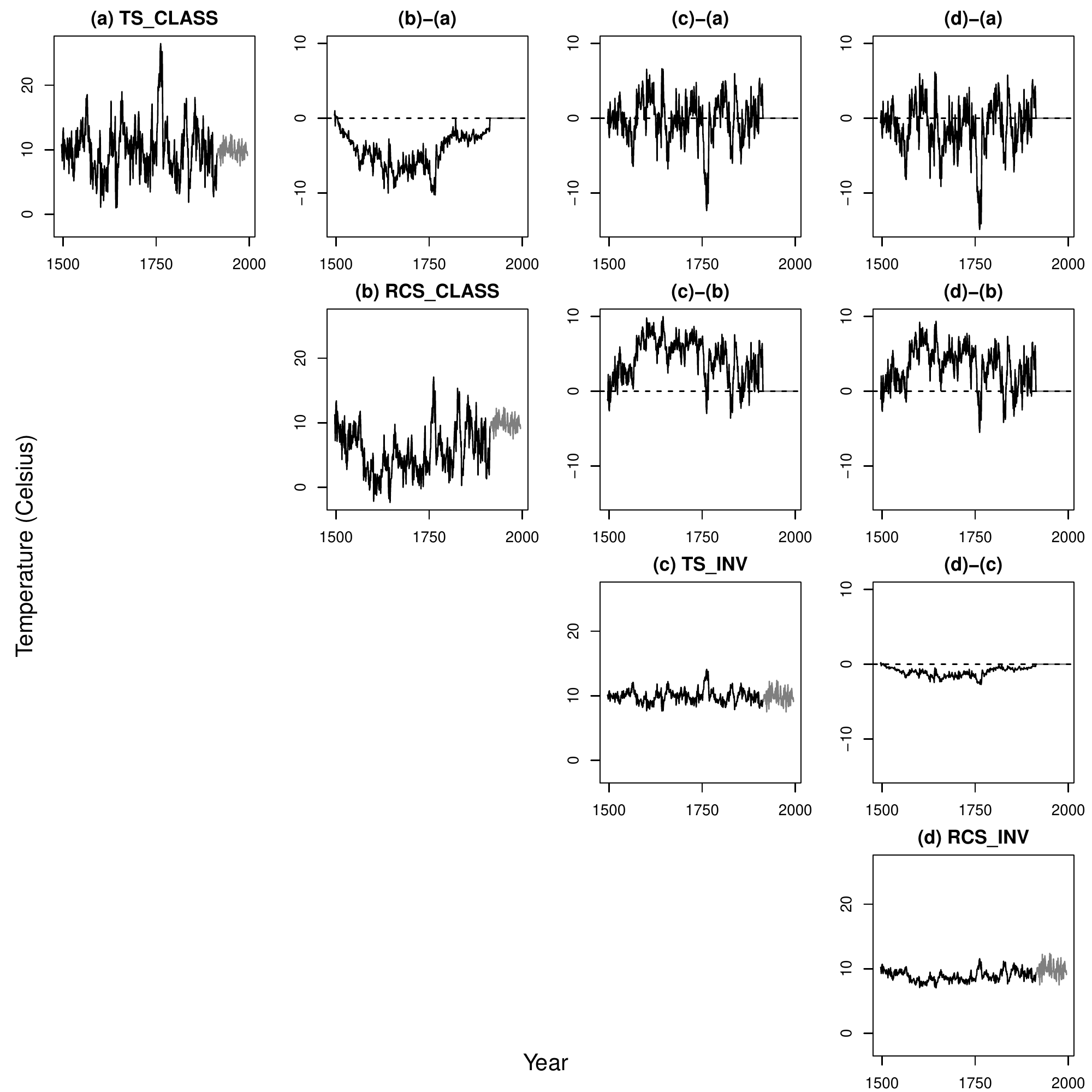}
   \caption{Predictions for different (i) standardization approaches and (ii) calibration procedures.  In the diagonal plots, the black lines are the predicted reconstruction and the gray lines are the observed data.  The off diagonal plots give the difference between the predictions in the corresponding diagonal plots.}
   \label{fig:tradcomp}
\end{figure}

The observed temperatures are a highly significant predictor of the chronologies found using both TS and RCS standardizations (both have $p$-value $< 0.00001$).  
The resulting temperature predictions exhibit differences we would expect given the properties of the procedures outlined in sections \ref{sect:typical}, \ref{sect:RCS}, \ref{sec:StatisticalCalibration} and S2 (Figure \ref{fig:tradcomp}).
The predictions from inverse calibration are shrunk toward the distribution of observed temperatures relative to the predictions from classical calibration (Figure \ref{fig:tradcomp}(a) vs \ref{fig:tradcomp}(c) or Figure \ref{fig:tradcomp}(b) vs \ref{fig:tradcomp}(d)).  
The predictions from TS are centered around the mean of the observed temperature (Figure \ref{fig:tradcomp}(a) and \ref{fig:tradcomp}(c)) and do not show the lower temperatures from 1600--1800 evident in the predictions from RCS (Figure \ref{fig:tradcomp}(b) and Figure \ref{fig:tradcomp}(d)).
This is evidence of the segment length curse.  Due to the short segment 
lengths, TS effectively removes any low-frequency temperature signals with period over $200$ years of length.

Almost all climate reconstructions use inverse calibration or related approaches.  For example, the description of calibration in \cite{Cook1990} only considers regressing climate ($\bm{x}$) on the tree-ring chronologies ($\bm{z}$).  In the example above, reconstructions using classical calibration result in scientifically implausible predictions, with some below \SI{0}{\celsius} (Figure \ref{fig:tradcomp}).  Judging the reconstructions and choice of procedure based on whether the predictions are plausible seems appealing.  However, the statistical properties of the calibration procedures (section \ref{sec:StatisticalCalibration} and S2) provide important context to a discussion on choice of method.  One such property is that classical calibration is preferred in terms of MSE over inverse calibration when the true value $x^{mis}_{t}$ is far from the mean $\bm x^{obs}$ value.  The implication is that classical calibration should be preferred when inferring unusual (in a historical context) climate values.  

The conflict between statistical and non-statistical arguments is concerning. Many reconstructions are undertaken in the belief that climate conditions are likely to change with time, in spite of the assumed model. 
Thus, a likely explanation for the predictions from inverse calibration appearing more plausible that those from classical calibration is the insensitivity 
to misspecification of the model for $\bm y|\bm x$ arising from the strong model for $\bm x$ that shrinks the predictions toward the observed climate values.  This shrinkage may suggest plausibility, but at the expense of masking the amplitude of temporal variation in climate.  Likewise, scientific implausibility evident in the predictions from classical calibration suggests that the underlying assumptions are inappropriate.

Instead of choosing between procedures, we prefer to 
consider a modeling framework that includes these two procedures as special cases.  This provides flexibility to allow other assumptions that may be more appropriate for the observed data.
Such a model should include the joint effects of age and climate to allow us to propagate uncertainty from the standardization through to the reconstruction.  A benefit of this approach is that it allows us to overcome the segment length curse (more details are provided in S5).


We will focus on two related assumptions that we believe are critical to the standardization and reconstruction processes.  The first is the choice of the model for climate ($\bm x$).  The differences evident in Figure \ref{fig:tradcomp} suggest that the probability model specified for the growing season temperatures (as implied by the different analyses) may have a large effect on predictions.  
The second focus is on the description of how the ring widths depend on growing season temperature.  

\section{Model-based description}\label{sect:mod1}
We initially present two models that are model-based analogues of the current TS and RCS standardization approaches.  That is, we specify probability models that include assumptions that are consistent with those used in the multi-step TS and RCS procedures. 

\subsection{TS model-based analogue}
We model individual tree-ring growth measurements with an effect due to age.  The model is linear in age on the log scale (equivalent to a negative exponential model on the original scale as used in the previous section) to account for  the assumption of multiplicative errors implied by the standardization processes. 
The model is
\begin{linenomath*}
\begin{align}\label{eq:simplemod}
  \log(y_{it}) &\stackrel{\text{ind.}}{\sim} \mathcal{N}(\beta_{0i} + \beta_{1i}a_{it} + \eta_{t},\sigma^{2}_{y}),~~i=1,\ldots,k,~~t=f_{i},\ldots,l_{i},\\\label{eq:etamod}
  \eta_{t} &\stackrel{\text{ind.}}{\sim} \mathcal{N}(\beta_2 x_{t},\sigma^{2}_{\eta}),~~t=1,\ldots,n,\\\label{eq:simplexmod}
  x_{t} &\stackrel{\text{iid}}{\sim} \mathcal{N}(\mu_{x},\sigma^{2}_{x}),~~t=1,\ldots,n,
\end{align}
\end{linenomath*}
where ind. denotes an independent sequence of random variables.  We further assume that $\beta_{0i}$ and $\beta_{1i}$ are drawn from common distributions, 
\begin{linenomath*}
\begin{equation}\label{eq:betamod}
  \beta_{hi} \stackrel{\text{ind.}}{\sim} \mathcal{N}(\mu_{\beta_{h}},\sigma^{2}_{\beta_{h}}),~~h = 0,1,~~i=1,\ldots,k.
\end{equation}
\end{linenomath*}
An important inclusion is the term $\eta_t$, assumed to be common between trees.  This term reflects that TS uses an average residual on the log scale at each time in order to find the chronology.  The difference here is that we do not simply estimate a value for $\eta_{t}$ that we use in a subsequent analysis.  Rather, we consider a hierarchical model for $\eta_{t}$ in terms of the climate variables $x_{t}$.  That is, the ring widths are modeled simulatenously in terms of both age and climate.  As described in S5, it is this step that links standardization and reconstruction and allows us to overcome the segment length curse. 
The final term we specify is the model for $x_t$.  As with the implied model for inverse calibration, the first model we consider treats $x_t$ as normally distributed with a constant mean through time.  We refer to the model given by (\ref{eq:simplemod}) - (\ref{eq:betamod}) as \mtscon.  

To aid in our understanding of the model (and how it compares to models we consider later) we can express \mtscon as
\begin{linenomath*}
\begin{equation}\label{eq:simplemodalt}
\log(y_{it}) \stackrel{\text{ind.}}{\sim} \mathcal{N}(\beta^{\prime}_{0i} + \beta_{1i}a_{it} + \beta_{2}x^{\prime}_{t} + \eta^{\prime}_{t},
\sigma^{2}_{y}),~~i=1,\ldots,k,~~t=f_{i},\ldots,l_{i},
\end{equation}
\end{linenomath*}
where the parameters are
\begin{linenomath*}
\begin{alignat*}{3}
\beta^{\prime}_{0i} &\equiv \beta_{0i} + \beta_{2}\mu_{x} &&\text{ such that } \beta^{\prime}_{0i} && \stackrel{\text{iid}}{\sim} \mathcal{N}(\mu_{\beta_0}+\beta_2\mu_x,\sigma^{2}_{\beta_{0}}),\\
x^{\prime}_{t} &\equiv x_{t}-\mu_{x} &&\text{ such that } x^{\prime}_{t} && \stackrel{\text{iid}}{\sim} \mathcal{N}(0,\sigma^{2}_{x}),\\
\eta^{\prime}_{t} &\equiv \eta_t - \beta_2 x_{t} && \text{ such that } \eta^{\prime}_{t}   &&\stackrel{\text{iid}}{\sim} \mathcal{N}(0,\sigma^{2}_{\eta}).
\end{alignat*}
\end{linenomath*}
That is, an alternative way of considering \mtscon is that the model for $\log(y_{it})$ has three features: (i) a linear aging term, (ii) a term that describes the effect of the climate relative to its mean, and (iii) a zero-mean yearly error term that is common between the trees.  
The year to year variability that is common between trees and is not accounted for by the linear aging term is then shared between $x^{\prime}_{t}$ and $\eta^{\prime}_{t}$.  The corresponding variances $\sigma^{2}_{x}$ and $\sigma^{2}_{\eta}$ are identifiable due to $x^{obs}_{m+1},\ldots,x^{obs}_{n}$ being observed.


\subsection{RCS model-based analogue}
To include assumptions consistent with RCS we need only change the term (\ref{eq:simplemod}).  An assumption of RCS was that all trees of the same age have the same growth rate.  We include this assumption by replacing (\ref{eq:simplemod}) with
\[
  \log(y_{it}) \stackrel{\text{ind.}}{\sim} \mathcal{N}(\zeta_{a_{it}} + \eta_{t},\sigma^{2}_{y}),~~i=1,\ldots,k,~~t=f_{i},\ldots,l_{i},
\]
where $\zeta_{j}$ is the growth rate for all trees aged $j$ years.  Each $\zeta_j$ is relatively uninformed by the data, as it is informed by at most $k$ observations (and usually far fewer). 
One possible solution is to specify a hierarchical model for $\bm \zeta$ that reduces this variability and allows the values to change smoothly over time.  Another possibility, that we implement here, is to bin the ages (we use 10 yearly bins) so that $\zeta_{1}$ is the growth rate for all trees aged $1$ to $10$ years, $\zeta_{2}$ is the growth rate for all trees aged $11$ to $20$ years, etc, 
\begin{equation}\label{eq:simplemodrcs}
  \log(y_{it}) \stackrel{\text{ind.}}{\sim} \mathcal{N}(\zeta_{\lceil a_{it}/b\rceil} + \eta_{t},\sigma^{2}_{y}),~~i=1,\ldots,k,~~t=f_{i},\ldots,l_{i},
\end{equation}
where $b=10$ and $\lceil x \rceil$ is the smallest integer value greater than or equal to $x$.  We refer to the model given by (\ref{eq:simplemodrcs}), (\ref{eq:etamod}) and (\ref{eq:simplexmod}) as \mrcscon.

%

\subsection{Model fitting}
To fit \mtscon and \mrcscon to the Tornetr\"{a}sk data we use Bayesian inference and Markov chain Monte Carlo (MCMC) implemented in JAGS \cite[]{Plummer2003}.   Prior distributions for all parameters and details of the algorithm are specified in S6.  This approach differs from that taken by \cite{Haslett2006} who did not use the full likelihood for inference.  Instead they updated the parameters in the model for $\bm y|\bm x$ using only the observed data ($\bm y^{obs}$ and $\bm x^{obs}$) and then used these values to update $\bm x^{mis}$.  Other approaches to inference are also possible for fitting the models we have described, e.g. maximum likelihood estimation could be implemented via the EM algorithm \cite[]{Dempster1977}.

\begin{figure}[!htbp]
 \centering
   \includegraphics[width =\textwidth]{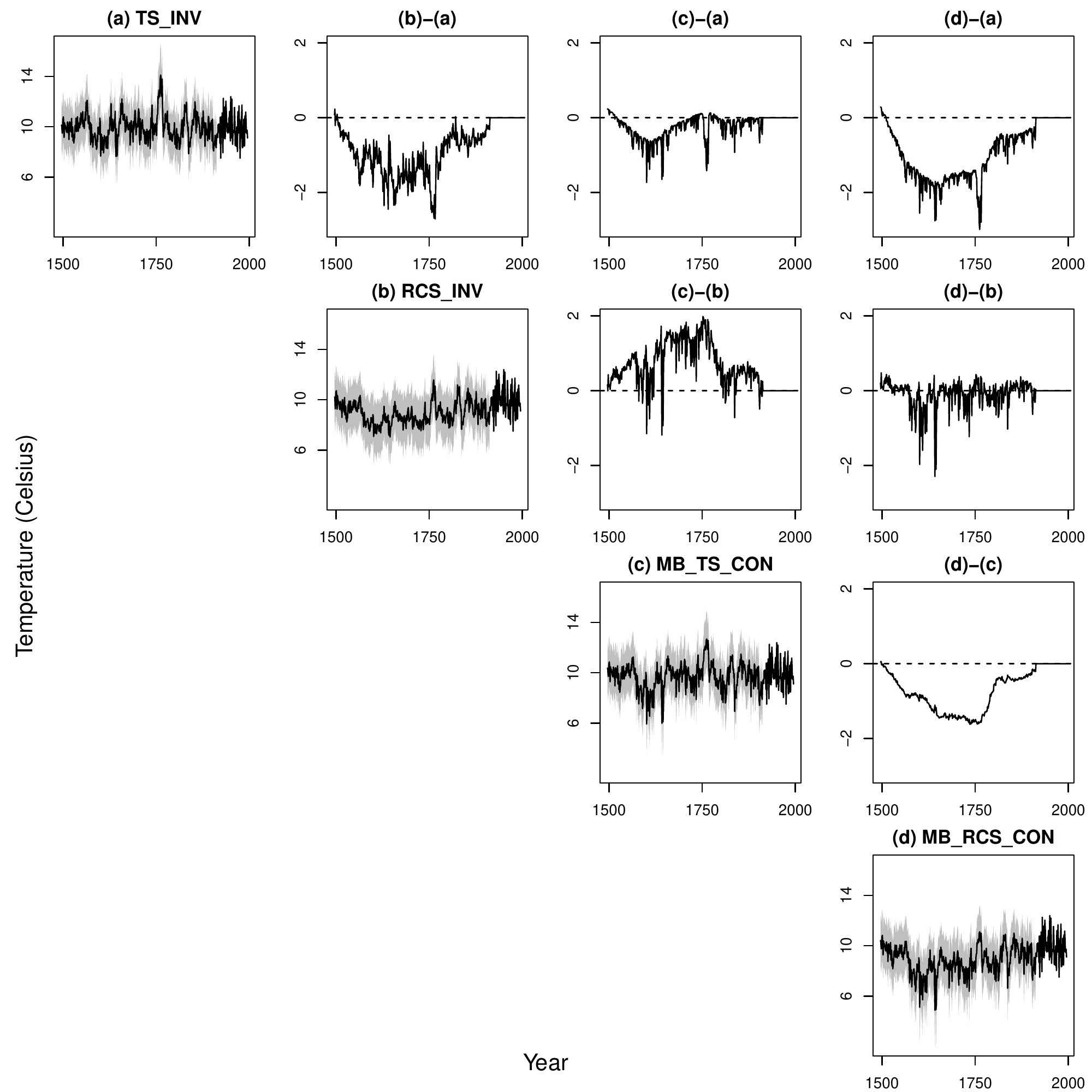}
    \caption{Predictions from traditional approaches and the model-based analogues.  In the diagonal plots, the black lines are either the predicted $\hat x^{mis}_{t}$ or the median of the posterior distribution for $x^{mis}_{t}$ and the gray areas are 95\% prediction/credible intervals.  The off diagonal plots give the difference between the predictions in the corresponding diagonal plots.}
    \label{fig:simpcomp}
\end{figure}

The reconstructions using multi-step procedures (\tsinv and \rcsinv) and their corresponding model-based analogues (\mtscon and \mrcscon) lead to similar overall patterns of variability in the predictions (Figure \ref{fig:simpcomp}).  The main difference between the predictions appears to be the magnitude of warming/cooling.  The actual time periods of relative warming and cooling are largely in common between all approaches.


The difference in reconstructions between assuming TS and RCS assumptions appears remarkably similar within a given modeling approach; i.e. compare Figure \ref{fig:simpcomp}(b)-(a) to \ref{fig:simpcomp}(d)-(c).  The RCS assumptions lead to a reconstruction with a pronounced period of cool temperatures from 1600 through 1750, whereas the TS assumptions favor a period of cooling from 1600--1650 followed by a return to stable temperature from 1650--1750.   

There are also differences that depend on the modeling approach.  In general, the model-based approach has resulted in predictions where the coolest periods (e.g. 1600 -- 1650) are colder than with the corresponding multi-step approach; i.e. \ref{fig:simpcomp}(d)-(b) for RCS and \ref{fig:simpcomp}(c)-(a) for TS.    

\begin{figure}[!htbp]
  \centering
    \includegraphics[width =\textwidth]{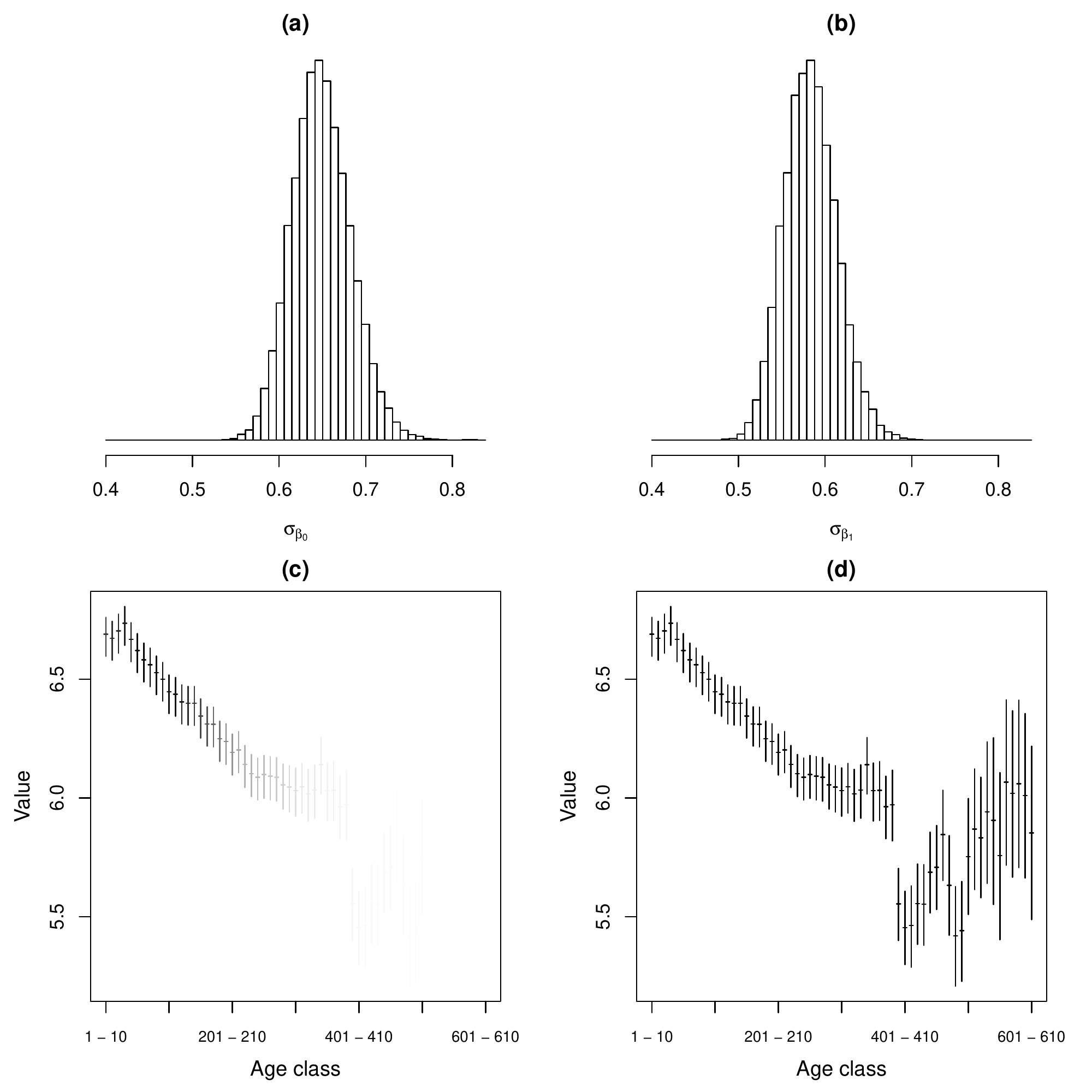}
     \caption{Various summaries from model-based TS and RCS implementations.  Estimates of the posterior density of $\sigma_{\beta_{0}}$ (a) and $\sigma_{\beta_{1}}$ (b) in \mtscon.  95\% central credible intervals for  $\zeta_{j}$ from \mrcscon are shown in (c) and (d), with the horizontal line denoting the posterior median.  In (c) the lines are shaded relative to the number of trees with observations in that age class. }
     \label{fig:rcstsass}
\end{figure}

Examination of various quantities in both the TS and RCS models can help us determine the adequacy of the underlying assumptions in the alternate model.  If the ring growth parameters $\beta_{0i}$ and $\beta_{1i}$ in \mtscon are similar for every tree in the sample then this supports the assumption made in RCS that growth is constant between individuals of the same age.  A simple approach is to examine the posterior densities for $\sigma_{\beta_0}$ and $\sigma_{\beta_1}$ to determine whether there is substantial mass near $0$.  For both parameters there is no evidence of mass near 0 (Figures \ref{fig:rcstsass}(a) and \ref{fig:rcstsass}(b)), suggesting that the constant growth assumption in RCS may not be appropriate.  Likewise, if the growth parameters $\zeta_{j}$ in \mrcscon show a close to linear trend with age this supports the assumption in TS about how ring width growth varies with age.  While far from definitive, the $\zeta_{j}$ values suggest that assuming linear growth on the log scale may be a reasonable assumption (Figures \ref{fig:rcstsass}(c) and \ref{fig:rcstsass}(d)), especially for the earlier age classes.  While the values for older age classes do depart from the general trend, the number of trees in each age class is relatively small.  The age classes with the most trees are $61-70$ and $71-80$ with $226$ trees.  In contrast, age class $301 - 310$ has 29 trees, $401 - 410$ has seven trees, $501 - 510$ has three trees and all age classes $541-550$ and older have only one tree.

\subsection{Extensions to the calibration model}
The comparison above examined the predictions from the joint model-based approach and those from a multi-step procedure.  A default choice was assumed for the model for temperature in all cases.  The choice was made so that the model-based approaches and multi-step procedures incorporated the same assumptions and were comparable.  We now investigate the effect this default choice has on the resulting reconstruction.  For this we only consider the ring width model in (\ref{eq:simplemod}) that uses TS and do not consider RCS any further.

The specific choice of model in (\ref{eq:simplexmod}) 
seems overly restrictive and it is doubtful that anyone interested in modeling long-term temperature believes that this is a reasonable hypothesis.  One possible alternative is to treat $x_{1},\ldots,x_{n}$ as if they were distinct parameters. 
Such an assumption is implied by classical calibration as described in S2.  However, it seems reasonable to assume that there will be some relationship between the year to year values of growing season temperature (or many climate variables).  This suggests any possible model should include structure in the pattern of changes in $\bm x$ over time. 
A model for $\bm x$ that includes some sort of smoothing thus seems desirable. One such example of this is given by \cite{Haslett2006} who looked at variance structures on random-walk models for the climate variable.

Our approach here is to model $\bm x$ hierarchically using a flexible family of distributions that admits a wide choice of scientific hypotheses about temporal changes in $\bm x$.  We do this by modeling $x_{t}$ as normally distributed, but allow the mean to vary according to a smooth function.  We use a cubic B-spline.  Now (\ref{eq:simplexmod}) is expressed as
\begin{linenomath*}
\begin{align}\label{eq:xmodext}
 x_{t} &\stackrel{\text{ind.}}{\sim} \mathcal{N}(\alpha_{t},\sigma^{2}_{x}), ~~t=1,\ldots,n,\\\nonumber
 \alpha_{t}&=\textstyle\sum_{h=1}^{H}\gamma_{h}B_{h}(x_{t}), ~~t=1,\ldots,n,
\end{align}
\end{linenomath*}
where $\alpha_{t}$ is the year-specific mean of the climate, and $B_{h}(x_{t})$ is the $h$th B-spline basis function \cite[see e.g.][]{Hastie2009}. 
We place a knot every $25$ years and include a hierarchical model on $\gamma_{h}$ to penalize the spline \cite[]{Eilers1996},
\begin{linenomath*}
\begin{equation}\label{eq:gammamod}
 \gamma_{h} \stackrel{\text{iid}}{\sim} \mathcal{N}(\mu_{\gamma},\sigma^{2}_{\gamma}),~~h=1,\ldots,H.
\end{equation}
\end{linenomath*}
If $\sigma^{2}_{\gamma}=0$ then the model for $\bm x$ in (\ref{eq:xmodext}) collapses to the original model for $\bm x$ in (\ref{eq:simplexmod}) with a constant mean through time.
We refer to the model given by (\ref{eq:simplemod}), (\ref{eq:etamod}), (\ref{eq:xmodext}), (\ref{eq:gammamod}) and (\ref{eq:betamod}) as \mtsspl.

As before, we can express \mtsspl as
\begin{linenomath*}
\begin{equation}\label{eq:extmodalt}
\log(y_{it}) \stackrel{\text{ind.}}{\sim} \mathcal{N}(\beta_{0i} + \beta_{1i}a_{it} + \beta_{2}\alpha_{t} + \beta_{2}x^{\prime}_{t} + \eta^{\prime}_{t},
\sigma^{2}_{y}),~~i=1,\ldots,k,~~t=f_{i},\ldots,l_{i},
\end{equation}
\end{linenomath*}
where the parameters are 
\begin{linenomath*}
\begin{alignat*}{3}
x^{\prime}_{t} &\equiv x_{t}-\alpha_{t} &&\text{ such that } x^{\prime}_{t} && \stackrel{\text{iid}}{\sim} \mathcal{N}(0,\sigma^{2}_{x}),\\
\eta^{\prime}_{t} &\equiv \eta_t - \beta_2 x_{t} && \text{ such that } \eta^{\prime}_{t}   &&\stackrel{\text{iid}}{\sim} \mathcal{N}(0,\sigma^{2}_{\eta}).
\end{alignat*}
\end{linenomath*}
The models (\ref{eq:simplemodalt}) and (\ref{eq:extmodalt}) are similar: both include a term that describes the effect of the climate relative to its mean and a zero-mean yearly error term that is common between trees. The difference is the inclusion of the term $\beta_{2}\alpha_{t}$ in (\ref{eq:extmodalt}).  That is, we can think of \mtsspl as extending the standardization model in \mtscon to include additional variation that is (i) common between trees, (ii) smooth through time, and (iii) directly related to the mean of the climate process.

We fit \mtsspl using MCMC, implemented in JAGS with details in S7.   A feature of this model is the presence of two modes in the likelihood.  We discuss the implications in S8.

\begin{figure}[!htbp]
 \centering
  \includegraphics[width =\textwidth]{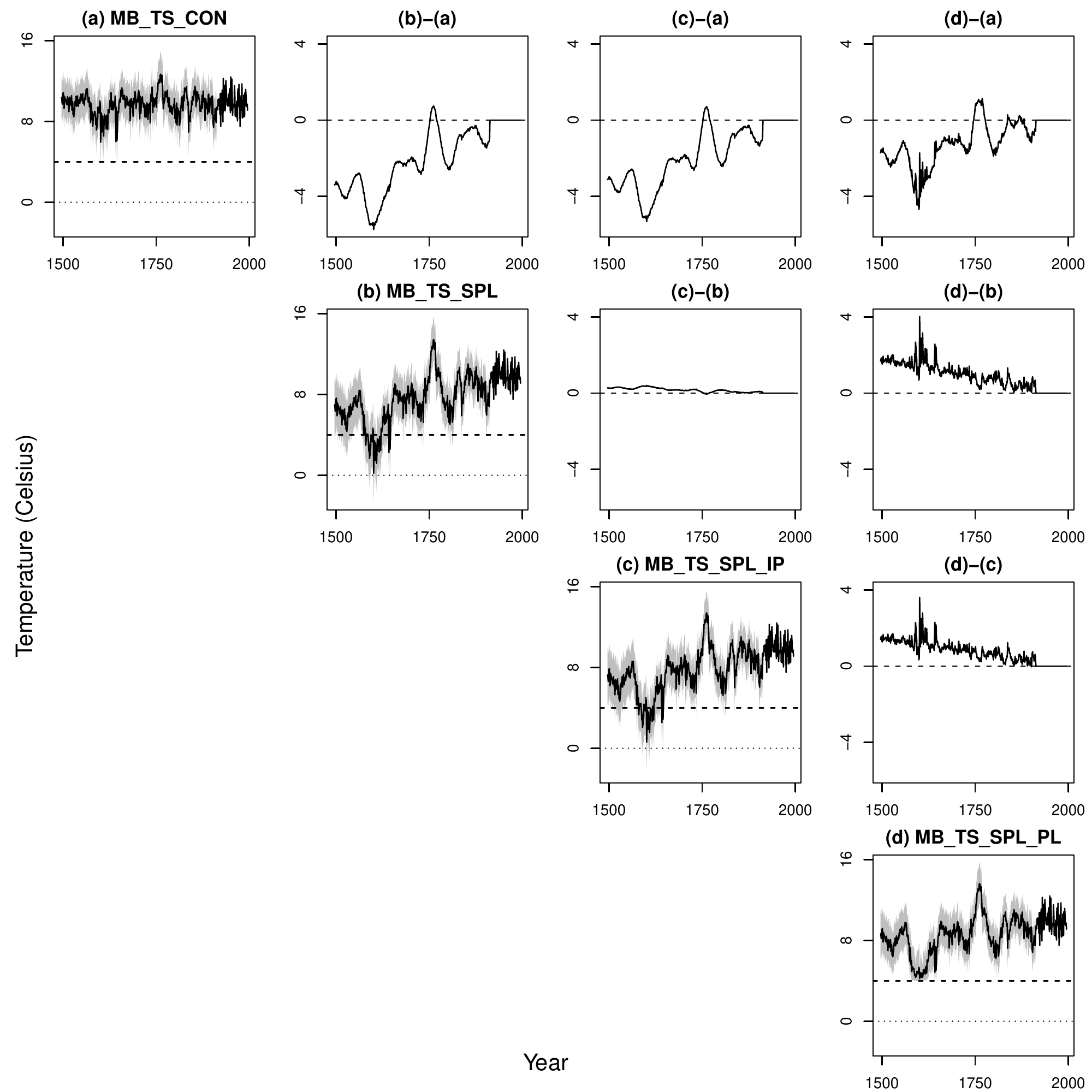}
  \caption{Predictions when making different assumptions in the model-based approaches.   In the diagonal plots, the black lines are the median of the posterior distribution for $x^{mis}_{t}$ and the gray areas are 95\% credible intervals.  The off diagonal plots give the difference between the predictions in the corresponding diagonal plots. In all plots the horizontal dashed line is at \SI{4}{\celsius} and the horizontal dotted line is at \SI{0}{\celsius}.}
  \label{fig:mbcomp}
\end{figure}

Changing the model for $\bm x$ has a substantial impact on the reconstructed temperatures (Figures \ref{fig:mbcomp}(a) and \ref{fig:mbcomp}(b)).  The amplitude of the predictions varies considerably between \mtscon and \mtsspl.  There is a difference of more than \SI{4}{\celsius} between the most extreme prediction made in the two models (in the period $1600-1650$).  
Given the sensitivity to the method of calibration 
shown in Figure \ref{fig:tradcomp}, it is unsurprising that the amplitudes change upon a generalization of the model for temperature.  Here we have relaxed the assumption of a constant mean for temperature which has increased the magnitude of the predictions.

The more flexible model for $\bm x$ has also led to a change in the overall mean of the predictions.  In particular, Figure \ref{fig:mbcomp}(b) suggests an increasing trend in temperatures throughout the entire series, apparent when comparing the two reconstructions in Figure \ref{fig:mbcomp}(b)-(a).  Such a result supports the notion that the simultaneous modeling of ring-width observations in terms of both climatic and non-climatic variables enables us to reconstruct low-frequency variability in the climate series. It is only once we relax the default model for $\bm x$ that such a trend is evident.  The information about the trend appears insufficient to overwhelm a model that assumes a constant mean for $\bm x$ through time, i.e. Figure \ref{fig:mbcomp}(a).

\subsection{Incorporating additional information}
The change in model for $\bm x$ has lead to predictions of temperature that are not scientifically reasonable.  There is a prolonged period where the temperature series is too cold for the trees to survive based on ecophysiological research \cite[]{Korner2004,Korner2008}.  A plausible lower bound for growing season temperature is around \SI{4}{\celsius} or \SI{5}{\celsius}.

There are many ways that additional information such as this can be incorporated into an analysis.  We discuss two such possibilities.  These analyses are intended to illustrate how such information can be included and the effect this has on the resulting predictions.  We have chosen the models and priors for the problem at hand, but we recognize that (i) many other possibilities exist, and (ii) some may disagree with the choices/approaches we have taken.  Even in these cases, we believe that exploring how the assumptions influence the predictions is instructive and aids in understanding of the data and model.

The first approach is to consider a model with informative or weakly-informative priors for all parameters.  
It is difficult to assign informative priors for the parameters that describe ring width.   Factors such as historical site condition play an important role and cannot be assessed until the tree-rings have been inspected during the crossdating procedure.
Therefore, we place heavy-tailed weakly-informative  independent prior distributions on the parameters $\mu_{\beta_{0}}$, $\sigma_{\beta_{0}}$, $\mu_{\beta_{1}}$, $\sigma_{\beta_{1}}$, $\sigma_{y}$, $\sigma_{\eta}$ and $\beta_{2}$. 

We specify informative priors on the parameters related to the model for $\bm x$.  
We place independent prior distributions on $\sigma_{x}$, $\mu_{\gamma}$ and $\sigma_{\gamma}$ that lead to implied prior probabilities  $\Pr(\SI{8}{\celsius} < x_{t} < \SI{12}{\celsius}) = 0.81$, $\Pr(x > \SI{6}{\celsius}) = 0.97$ and $\Pr(x > \SI{4}{\celsius}) = 0.99$.  Additional details are given in S9.  We refer to the model given by (\ref{eq:simplemod}), (\ref{eq:etamod}), (\ref{eq:betamod}), (\ref{eq:xmodext}) and (\ref{eq:gammamod}) with the priors specified in S9 as \mtssplsoft.

A second approach is to change the model for how ring widths are influenced by temperature, similar to that given in \cite{Tolwinski-Ward2011}.  We assume that there is a linear effect of temperature on log ring width growth between two temperature bounds that we refer to as $x_{\min}$ and $x_{\max}$.  Below $x_{\min}$ the tree does not grow (as we are on the log-scale this corresponds to a value of $-\infty$).  At $x_{\max}$ the growth rate is assumed optimal and all temperatures above $x_{\max}$ attain this optimal growth rate. 
We include this in the model by changing (\ref{eq:etamod}) to
\begin{linenomath*}
\begin{align}\label{eq:etainfo}
\eta_{t} &\stackrel{\text{ind.}}{\sim} \mathcal{N}(\mu_{t},\sigma^{2}_{\eta}),~~t=1,\ldots,n,\text{ where,}\\\nonumber
\mu_{t} &= \left\{\begin{array}{ll} 
-\infty & \text{if $x_{t} < x_{\min}$}\\\nonumber
\beta_{2}x_{t} & \text{if $x_{\min} \leq x_{t} \leq x_{\max}$}\\\nonumber
\beta_{2}x_{\max} & \text{if $x_{t} > x_{\max}$}\end{array}
 \right.
\end{align}
\end{linenomath*}
and ensuring the prior density for $\beta_{2}$ has positive support.
The values $x_{\min}$ and $x_{\max}$ are set by the user (we use \SI{4}{\celsius} and \SI{20}{\celsius} respectively).    
As this model is made up of piecewise linear growth increments, we refer to the model given by (\ref{eq:simplemod}), (\ref{eq:betamod}), (\ref{eq:xmodext}), (\ref{eq:gammamod}) and (\ref{eq:etainfo}) as \mtssplhard.  

For the Tornetr\"{a}sk data, the observed ring widths $\bm y^{obs}$ are all strictly positive.  Thus values $x_{t} < x_{\min}$ are inconsistent with the observed data for model \mtssplhard.  This is a strong constraint, but can be relaxed if desired (e.g. we could assume that when the temperature is below $x_{\min}$ the expected growth due to temperature is small, but non-zero).

We fit \mtssplhard and \mtssplsoft using MCMC, implemented in JAGS.  The MCMC details are practically equivalent to those given in 
S7 except for the changes to prior distributions.


The informative prior has little effect on the predictions (Figure \ref{fig:mbcomp}(c)-(b)).  Despite a prior probability of $0.01$ that $x_{t} < \SI{4}{\celsius}$, many of the observations between 1600 -- 1650 are below this value.  The data have overwhelmed the prior.  The prior densities we have specified have not been strong enough to ensure reasonable predictions from an ecophysiological perspective.

The model \mtssplhard has had the expected effect on the prediction of temperature (Figure \ref{fig:mbcomp}(d)) with no predictions below the value $x_{\min} = 4$.  Comparing the predictions from \mtsspl and \mtssplhard reveals interesting differences.  Not only does the amplitude change (to within the ecophysiological limits defined by the modified model) but the overall temperature trend changes.  There is still an increasing trend through time (Figure \ref{fig:mbcomp}(d)) but the trend is smaller than in model \mtsspl (Figure \ref{fig:mbcomp}(d)-(b)).

%% file: modcheck.tex
\section{Model checking}\label{sect:modcheck}
To explore predictive performance of the models, we alternatively hold-out the first and second halves of the observed temperature data.  This approach is commonly used in dendrochronology \cite[e.g.][]{Cook1990} as the primary goal of such analyses is prediction outside of the observed sample.  However, we note that any approach to assess the predictive performance is limited to the range of observed temperature values.
Furthermore, since the hold-out blocks remove half of the observed temperatures used for model fitting, the resulting predictions will be correspondingly less certain.

\begin{figure}[!htbp]
  \centering
  \includegraphics[width =0.8\textwidth]{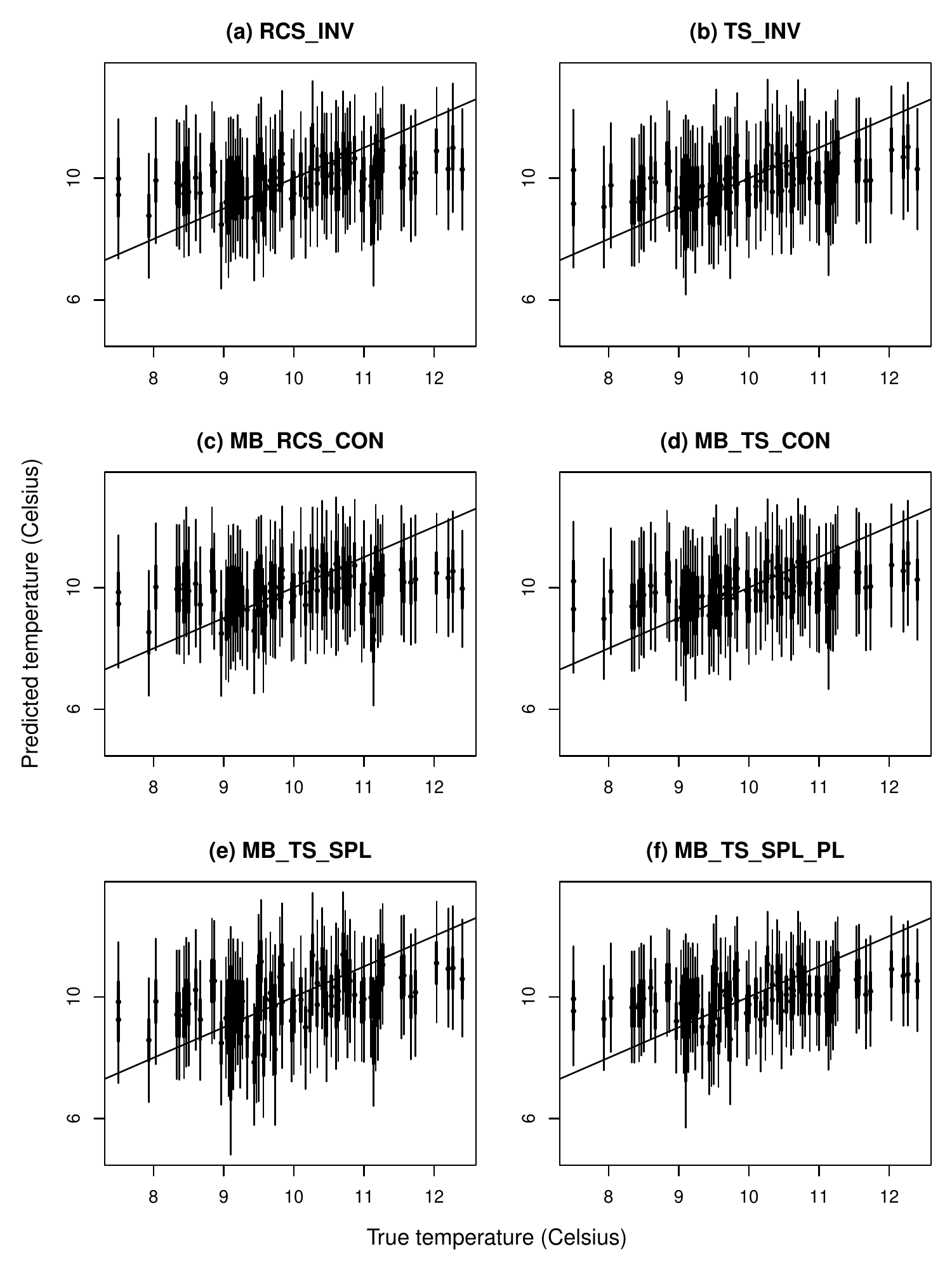}
  \caption{Predictions (and corresponding uncertainty intervals) plotted against the true values for the held-out observations.  
  In plots (a) and (b), the thin lines give a  95\% prediction interval, the thick lines are the 50\% prediction interval and the point is the prediction for the held-out values.  For (c) -- (f) the thin lines are 95\% posterior predictive credible intervals, the thicker lines are 50\% credible intervals and the point is the median of the posterior predictive distribution for the held-out values. }
  \label{fig:cv}
\end{figure}

There appears to be little difference between the various approaches in terms of predictive performance (Figure \ref{fig:cv}).  All approaches are likely to overestimate when the true  temperature value is below the observed mean and vice versa.  There appears to be little difference between the models with the smooth mean for temperature and those with constant mean.  This is somewhat surprising given the difference in predictions based on these models (the differences observed in Figure \ref{fig:mbcomp} also occur when modeling with the hold-out samples removed).  
Based on the values from the hold-out samples, all approaches (including those with scientifically implausible extremes) appear to be conservative at predicting values that are further from the observed mean.  RMSE values based on the hold-out are presented in Table 1 in the supplementary materials.
We have not presented model checks for \mtssplsoft as its predictions were nearly identical to those from model \mtsspl.

\begin{figure}[!htbp]
      \centering
      \includegraphics[width =0.8\textwidth]{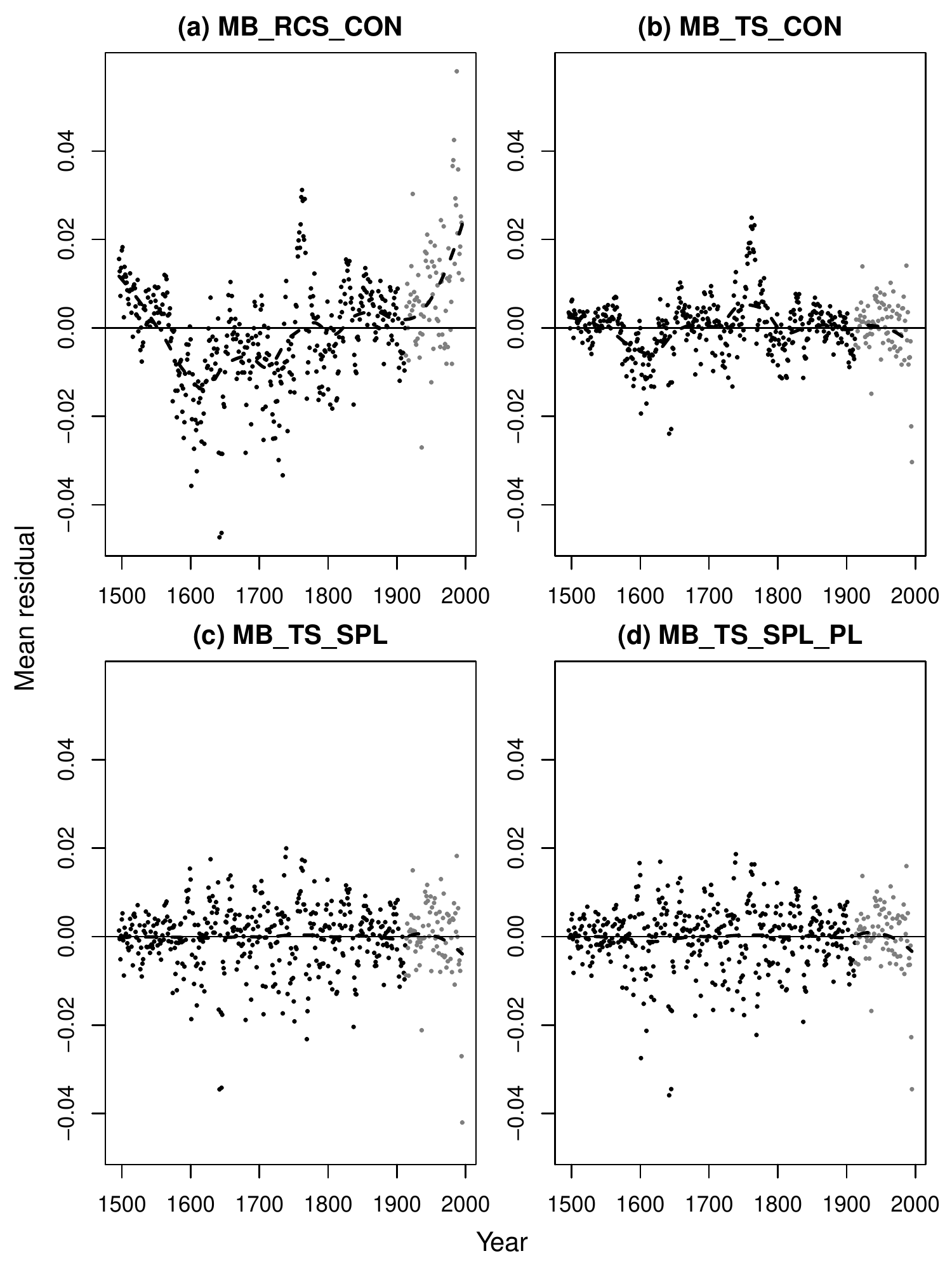}
      \caption{The average residual plotted against year.  The average residual in year $t$ is found by averaging the residuals (found at the posterior mean of the parameters) across all trees observed in year $t$.  The gray dots refer to years in which the temperature values were observed and black dots to years in which the climate values were not observed.  The dashed line is a loess fit.}
      \label{fig:resid}
\end{figure}

A key advantage of the model-based approach compared to the multi-step procedures is the ability to explore other measures of model fit.  One such approach is looking at the fit of the conditional model $\bm y|\bm x$.
We do this using averaged residuals.  That is, we find the residual
$\delta_{it} = y_{it}-\hat{y}_{it}$, where $\hat{y}_{it} = \bar{\beta}_{0i} + \bar{\beta}_{1i}a_{it} + \bar{\eta}_{t}$ for TS and $\hat{y}_{it} = \bar{\zeta}_{\lceil a_{it}/10\rceil} + \bar{\eta}_{t}$ for RCS, with a bar denoting the posterior mean of the unknown quantity.  We then find the average residual, $\tilde{\delta}_{t}$ by taking the mean of these residuals across all trees with observations in year $t$.  
For both models with constant mean for $x_t$, there is a pattern in the average residuals over time (Figures \ref{fig:resid}(a) and \ref{fig:resid}(b)).  
 No such pattern is obvious in the models that include the smooth mean for climate (Figures \ref{fig:resid}(c) and \ref{fig:resid}(d)).  However, these residual plots suggest that there may be non-constant variance through time.  This appears to be an issue for both the period of observed temperatures, as well as the periods where temperature was not measured.  

The RCS residuals appear more variable that the TS residuals.  
Few averaged residuals in the TS models exceed $0.02$ units (ring width on the log-scale), whereas the RCS model has many averaged residuals beyond $0.02$ units and some beyond $0.04$.  This is consistent with the posterior distribution for $\sigma_y$ under the different models.  The central $95\%$ credible interval for $\sigma_y$ is $(0.545,0.552)$ in \mrcscon  whereas the interval is $(0.362,0.366)$ in \mtscon.  We believe this is due to unmodeled variability; the RCS model assumes a common age-based growth rate.
 
%

Model \mtsspl allows mean temperature to evolve smoothly over time, but results in scientifically implausible predictions.  Despite the obvious shortcomings with the predictions from this model, we have presented the results and explored model checks.  These checks suggest that the fit and predictions of the model (for the hold-out samples) are as good or better than the other models we have presented that do have scientifically plausible predictions.  
This demonstrates that the data alone are insufficient to discriminate between the various models we have explored.  External information is needed to resolve the differences.   What is less clear is how that should be achieved.  We considered two approaches to include that information in the modeling process.  The first considered soft constraints given by informative priors.  The data overwhelmed the prior and the predictions remained implausible.  In the second approach, we included a hard constraint associated with a change of response function (describing how tree-rings respond to temperature).  The predictions changed considerably with the hard constraint ensuring plausible values (at least according to a minimum temperature).  Other changes to the response function that we have investigated (and not presented) suggest that the model for describing how climate (in our case temperature) relates to ring widths has a large impact on the resulting predictions while often leaving the fit of the 
observed data  unchanged.


We believe that the statements of uncertainty we present for the temperature predictions understate the true uncertainty as all of our predictions condition on an assumed model.  For these data we appear unable to distinguish between various assumptions, even when there are substantial differences in predictions.  As there is no model clearly favored by the data and many modeling assumptions that could be considered (some additional assumptions are mentioned in section \ref{sect:discuss}), the results suggest: (i) modeling decisions matter, and (ii) the uncertainty we should have in the predictions is greater than that expressed by any one model. 

One way to account for model uncertainty is to model-average uncertainty intervals for the unknown temperature values \cite[e.g.][]{Hoeting1999}.  We have not done this for two reasons.  First, it is difficult to calculate the marginal likelihoods required to find the Bayes factors or, equivalently, specify a trans-dimensional MCMC algorithm, such as  reversible jump algorithm \cite[]{Green1995} to explore joint model and parameter space.  Second, Bayes factors are known to be sensitive to priors placed on parameters even if the data overwhelm the priors within specific models. For these reasons we have instead 
examined how the model fit and predictions of missing temperature observations change with the model assumed.   




%% file: discuss.tex
\section{Discussion}\label{sect:discuss}
Inferring historical climate from ring-width data has many sources of uncertainty.  Instead of trying to remove (or ignore) these in a stepwise manner, conditioning on model output each time, we have built a model-based framework for including these sources.  
Care is needed when specifying alternative assumptions to ensure that the overall model continues to be identifiable.  From the TS model statement in (\ref{eq:extmodalt}) we can see that including a flexible aging function such as a polynomial regression or spline would lead to identifiability problems 
if the climate variable also has a flexible mean 
function.  However, for the RCS representation the constraint that all trees of the same age have the same growth response makes it is possible to have a flexible aging function and allow the climate variable to have a flexible mean.

We have explored changes to many of the model assumptions, including the description of ring-width growth (TS vs RCS), how the ring widths depend on climate (including using or not using additional information), and the model for climate.  We have shown that changes to these assumptions can lead to important differences in resulting predictions.  Moreover, the validity of many of these assumptions cannot be determined based solely on the data.  
Reconstructions found using chronologies formed by multi-step procedures and obtained using default calibration approaches may hide such sensitivity to underlying modeling assumptions.
We have restricted ourselves to one transformation on $\bm y$ that was chosen to provide a link between the traditional multi-step approaches and the model-based approach.  While breaking the link to the traditional multi-step approaches, other transformations of $\bm y$ could be used in place of the log transform.  
For example, similar results and conclusions are obtained for this data when using a square root transform instead of the log transform. 
Other transformations, such as a Box-Cox transformation, could also be used. 

The crossdating process (described in S1) can be thought of as obtaining an estimate of the overall chronology.  
There may be uncertainty in the estimates of ring-width dates \cite[]{Wigley1987}.  Any such unmodeled uncertainty will not propagate through to the reconstruction and will result in predictions of historical climate that appear more precise than they should be.  
Potential remedies include (i) specifying the crossdating process within a probabilistic statistical model, or (ii) having the dendrochronologist estimate uncertainty in the cross-dating process, i.e. giving various plausible dates for the series with corresponding estimates of certainty in that being the correct date (that sum to one).  
Inclusion of the crossdating process would be a difficult task.  Setting up such a model (option (i) above) would require research effort, because of the challenge of including all the information that is currently used in the human judgment involved in the visual aspect of crossdating through a microscope.   Moreover, there would be 
considerable computational difficulties involved in  including any dating uncertainty.  

A benefit of including the crossdating process within the statistical model 
is that it would avoid using the data twice.  Many crossdating procedures use the data, and the assumption that trees respond in a related way, to estimate the correct alignment of the cores.  Once these estimates are obtained, the data are then used again to predict historical climate, assuming that the relatedness in the tree response depends on the climatic variable of interest.

Another assumption we have not varied is that the growth increments have a common response to temperature.   Despite the trees existing in the same macro-environment, there may be fine scale local differences in the temperature, leading to a tree-specific response.  Extending the models in this way is ongoing research and allows the inclusion of more realistic assumptions at the expense of several practical difficulties.  Model fitting becomes difficult, with care required to ensure the model remains identifiable.  Our experience is that there is considerable sensitivity to the specification of the error structure in the model.  Furthermore, some extensions necessitated the use of Hamiltonian Monte Carlo algorithms that are difficult to implement; Gibbs/Metropolis approaches appear impractically slow.

The idea of a common modeling framework can be extended to multiple proxies 
\cite[]{Evans2013}.  Unfortunately, other proxy variables also include complexities that must be accounted for in the modeling.   
Despite these difficulties, including multiple proxies in a reconstruction can be of value, particularly when different proxy types reflect different aspects of the climate process \cite[for example, see][]{Li2010}.
In particular, the multiple proxies may provide information about the calibration curve across a range of climate values.  However, including observations from multiple proxies will not automatically lead to robust predictions. 
An advantage of including data from multiple proxies is that it allows for more expansive model checking.  That is, we may be able to use information from different proxies to check the appropriateness of different aspects of our assumed model.

There are many hierarchical extensions that we could consider.  These include data from different geographic locations with appropriate spatial models, such as the those considered by \cite{Tingley2010}.  
Other possibilities include models for climate that include covariates \cite[for example][]{Li2010} or that incorporate physical models.  The inclusion of such extensions provides two substantial future challenges.  The first is computational: 
developing scalable methodology that allows us to model the raw data across numerous locations without resorting to multi-step procedures.  The second  challenge is one of complexity and applicability: developing appropriate model checks for the assumptions that, as shown here, are critical to historical prediction of climate.  

\subsection{Message for the paleoclimate community}
We have demonstrated model-based approaches for tree-ring based reconstructions that are able to incorporate the assumptions of traditional approaches as special cases.  The modeling framework allows us to relax assumptions long used out of necessity, giving flexibility to our model choices.  Using the Scots pine data from Tornetr\"{a}sk we show how modeling choices matter.  
Alternative models fitting the data equally well can lead to substantially different predictions. 
These results do not necessarily mean that existing reconstructions are incorrect.  If the assumptions underlying the  reconstruction is a close approximation of reality, the resulting prediction and associated uncertainty will likely be appropriate (up to the problems associated with the two-step procedures used).  However, if we are unsure whether the assumptions are correct and there are other assumptions equally plausible \textit{a-priori}, we will have unrecognized  uncertainty in the predictions. 
We believe that such uncertainty should be acknowledged when using standardized data and default models.  

As an example consider the predictions from model \mtscon for Abisko, Sweden. 
If we believe the assumptions underlying model \mtscon then there is a 95\% probability that summer mean temperature in 1599 was between \SI{8.1}{\celsius} and \SI{12.0}{\celsius} as suggested by the central credible interval (Figure \ref{fig:mbcomp}(a)).  However, if we adopt the assumptions underlying model \mtssplhard we would believe that the summer mean temperature in 1599 may have been much colder than \SI{8.1}{\celsius} with a 95\% credible interval between \SI{4.1}{\celsius} and \SI{7.8}{\celsius}.  In practice, unless the data are able to discriminate between these assumptions (which they were not able to do here as shown in Section \ref{sect:modcheck}), there is more uncertainty about the summer mean temperature in 1599 than that found in any one model considered.
%
We believe that such model uncertainty needs to be recognized by the community as an important source of uncertainty associated with predictions of historical climate.  The use of default methods makes evaluation of such uncertainty difficult. 


%% file: acknow.tex
\section*{Acknowledgements}
This research was supported by NSF \#0934516.  We thank three annonymous referees and the Associate Editor for valuable comments that improved this manuscript.  Lamont-Doherty Contribution Number 0000.

%% file: figures.tex
\noindent \Large{\bf Tables}
\begin{table}[!htbp]
\begin{minipage}{\textwidth}
  \begin{center}
    \begin{tabularx}{\linewidth}{lX}
    \tsinv & Multi-stage approach using TS and inverse calibration\\
    \tsclass & Multi-stage approach using TS and classical calibration\\
    \rcsinv & Multi-stage approach using RCS and inverse calibration\\
    \rcsclass & Multi-stage approach using RCS and classical calibration\\
    \mtscon & Model-based approach using (\ref{eq:simplemod}) -- (\ref{eq:betamod})\\
    \mtsspl & Model-based approach using (\ref{eq:simplemod}), (\ref{eq:etamod}), (\ref{eq:betamod}), (\ref{eq:xmodext}) and (\ref{eq:gammamod})\\
    \mtssplhard & Model-based approach using (\ref{eq:simplemod}), (\ref{eq:betamod}), (\ref{eq:xmodext}), (\ref{eq:gammamod}) and (\ref{eq:etainfo})\\
    \mtssplsoft & Model \mtsspl with informative priors described in S9. \\        
    \mrcscon & Model-based approach using (\ref{eq:etamod}), (\ref{eq:simplexmod}) and (\ref{eq:simplemodrcs})\\
    \end{tabularx}   
\caption{A description of all reconstructions considered.} 
\label{tab:models}
  \end{center}
\end{minipage}
\end{table}

%% file: ClimRecon.bbl
\begin{thebibliography}{40}
\newcommand{\enquote}[1]{``#1''}
\expandafter\ifx\csname natexlab\endcsname\relax\def\natexlab#1{#1}\fi

\bibitem[{Briffa et~al.(1992)Briffa, Jones, Bartholin, Eckstein, Schweingruber,
  Karlen, Zetterberg, and Eronen}]{Briffa1992}
Briffa, K.~R., Jones, P.~D., Bartholin, T.~S., Eckstein, D., Schweingruber,
  F.~H., Karlen, W., Zetterberg, P., and Eronen, M. (1992),
  \enquote{Fennoscandian summers from \textrm{AD} 500: temperature changes on
  short and long timescales,} \textit{Climate Dynamics}, 7, 111--119.

\bibitem[{Briffa and Melvin(2011)}]{Briffa2011}
Briffa, K.~R. and Melvin, T.~M. (2011), \enquote{A closer look at regional
  curve standardisation of tree-ring records: justification of the need, a
  warning of some pitfalls, and suggested improvements in its application.} in
  \textit{Dendroclimatology: Progress and Prospects}, eds. Hughes, M.~K., Diaz,
  H.~F., and Swetnam, T.~W., Springer Verlag, pp. 113 -- 145.

\bibitem[{Briffa et~al.(2008)Briffa, Shishov, Melvin, Vaganov, Grudd,
  Hantemirov, Eronen, and Naurzbaev}]{Briffa2008}
Briffa, K.~R., Shishov, V.~V., Melvin, T.~M., Vaganov, E.~A., Grudd, H.,
  Hantemirov, R.~M., Eronen, M., and Naurzbaev, M.~M. (2008), \enquote{Trends
  in recent temperature and radial tree growth spanning 2000 years across
  northwest Eurasia,} \textit{Philosophical Transactions of the Royal Society
  B: Biological Sciences}, 363, 2269--2282.

\bibitem[{Christiansen(2011)}]{Christiansen2011}
Christiansen, B. (2011), \enquote{Reconstructing the {NH} mean temperature: can
  underestimation of trends and variability be avoided?} \textit{Journal of
  Climate}, 24, 674--692.

\bibitem[{Christiansen(2014)}]{Christiansen2014}
--- (2014), \enquote{Straight line fitting and predictions: on a marginal
  likelihood approach to linear regression and errors-in-variables models,}
  \textit{Journal of Climate}, 27, 2014--2031.

\bibitem[{Christiansen and Ljungqvist(2011)}]{Christiansen2011a}
Christiansen, B. and Ljungqvist, F.~C. (2011), \enquote{Reconstruction of the
  extratropical {NH} mean temperature over the last millennium with a method
  that preserves low-frequency variability,} \textit{Journal of Climate}, 24,
  6013--6034.

\bibitem[{Cook et~al.(1995)Cook, Briffa, Meko, Graybill, and
  Funkhouser}]{Cook1995}
Cook, E.~R., Briffa, K.~R., Meko, D.~M., Graybill, D.~A., and Funkhouser, G.
  (1995), \enquote{The `segment length curse' in long tree-ring chronology
  development for palaeclimatic studies,} \textit{The Holocene}, 5, 229--237.

\bibitem[{Cook and Kairiukstis(1990)}]{Cook1990}
Cook, E.~R. and Kairiukstis, L.~A. (eds.) (1990), \textit{Methods of
  Dendrochronology}, Kluwer Academic Publishers.

\bibitem[{Dempster et~al.(1977)Dempster, Laird, and Rubin}]{Dempster1977}
Dempster, A.~P., Laird, N.~M., and Rubin, D.~B. (1977), \enquote{Maximum
  likelihood from incomplete data via the {EM} algorithm,} \textit{Journal of
  the Royal Statistical Society B}, 39, 1--38.

\bibitem[{Eilers and Marx(1996)}]{Eilers1996}
Eilers, P. H.~C. and Marx, B.~D. (1996), \enquote{Flexible smoothing with
  \text{B}-splines and penalties,} \textit{Statistical Science}, 11, 89--121.

\bibitem[{Esper et~al.(2009)Esper, Krusic, Peters, and Frank}]{Esper2009}
Esper, J., Krusic, P.~J., Peters, K., and Frank, D. (2009),
  \enquote{Exploration of long-term growth changes using the tree-ring
  detrending program {``Spotty"},} \textit{Dendrochronologia}, 27, 75--82.

\bibitem[{Evans et~al.(2013)Evans, Tolwinski-Ward, Thompson, and
  Anchukaitis}]{Evans2013}
Evans, M.~N., Tolwinski-Ward, S.~E., Thompson, D.~M., and Anchukaitis, K.~J.
  (2013), \enquote{Applications of proxy system modeling in high resolution
  paleoclimatology,} \textit{Quaternary Science Reviews}, 76, 16--28.

\bibitem[{Fritts(1976)}]{Fritts1976}
Fritts, H.~C. (1976), \textit{Tree Rings and Climate}, Academic Press: London.

\bibitem[{Fritts and Swetnam(1989)}]{Fritts1989}
Fritts, H.~C. and Swetnam, T.~W. (1989), \enquote{Dendroecology: a tool for
  evaluating,} \textit{Advances in Ecological Research}, 19, 111.

\bibitem[{Green(1995)}]{Green1995}
Green, P.~J. (1995), \enquote{Reversible jump {M}arkov chain {M}onte {C}arlo
  computation and {B}ayesian model determination,} \textit{Biometrika}, 82, 711
  -- 732.

\bibitem[{Grudd et~al.(2002)Grudd, Briffa, Karlen, Bartholin, Jones, and
  Kromer}]{Grudd2002}
Grudd, H., Briffa, K.~R., Karlen, W., Bartholin, T.~S., Jones, P.~D., and
  Kromer, B. (2002), \enquote{A 7400-year tree-ring chronology in northern
  Swedish Lapland: natural climatic variability expressed on annual to
  millennial timescales,} \textit{The Holocene}, 12, 657--665.

\bibitem[{Haslett et~al.(2006)Haslett, Salter-Townshend, Wilson, Bhattacharya,
  Whiley, Allen, Huntley, and Mitchell}]{Haslett2006}
Haslett, J., Salter-Townshend, M., Wilson, S.~P., Bhattacharya, S., Whiley, M.,
  Allen, J. R.~M., Huntley, B., and Mitchell, F. J.~G. (2006),
  \enquote{Bayesian palaeoclimate reconstruction,} \textit{Journal of Royal
  Statistical Society A}, 169, 1--36.

\bibitem[{Hastie et~al.(2009)Hastie, Tibshirani, and Friedman}]{Hastie2009}
Hastie, T., Tibshirani, R., and Friedman, J. (2009), \textit{The Elements of
  Statistical Learning. Second Edition.}, Springer.

\bibitem[{Hoeting et~al.(1999)Hoeting, Madigan, Raftery, and
  Volinsky}]{Hoeting1999}
Hoeting, J.~A., Madigan, D., Raftery, A.~E., and Volinsky, C.~T. (1999),
  \enquote{Bayesian model averaging: a tutorial,} \textit{Statistical Science},
  14, 382--417.

\bibitem[{Jansen et~al.(2007)Jansen, Overpeck, Briffa, Duplessy, Joos,
  Masson-Delmotte, Olago, Otto-Bliesner, Peltier, Rahmstorf, Ramesh, Raynaud,
  Rind, Solomina, Villalba, and Zhang}]{Jansen2007}
Jansen, E., Overpeck, J., Briffa, K.~R., Duplessy, J.~C., Joos, F.,
  Masson-Delmotte, V., Olago, D., Otto-Bliesner, B., Peltier, W.~R., Rahmstorf,
  S., Ramesh, R., Raynaud, D., Rind, D., Solomina, O., Villalba, R., and Zhang,
  D. (2007), \enquote{Contribution of Working Group I to the Fourth Assessment
  Report of the Intergovernmental Panel on Climate Change,} in \textit{Climate
  Change 2007: The Physical Science Basis}, Cambridge University Press,
  Cambridge, United Kingdom and New York, NY, USA.

\bibitem[{Jones et~al.(2009)Jones, Briffa, Osborn, Lough, Van~Ommen, Vinther,
  Luterbacher, Wahl, Zwiers, Mann, Schmidt, Ammann, Buckley, Cobb, Esper,
  Goosse, Graham, Jansen, Kiefer, Kull, K\"{u}ttel, Mosley-Thompson, Overpeck,
  Riedwyl, Schulz, Tudhope, Villalba, Wanner, Wolff, and Xoplaki}]{Jones2009}
Jones, P.~D., Briffa, K.~R., Osborn, T.~J., Lough, J.~M., Van~Ommen, T.~D.,
  Vinther, B.~M., Luterbacher, J. W. E. R. Z. F.~W., Wahl, E.~R., Zwiers,
  F.~W., Mann, M.~E., Schmidt, G.~A., Ammann, C.~M., Buckley, B.~M., Cobb,
  K.~M., Esper, J., Goosse, H., Graham, N., Jansen, E., Kiefer, T., Kull, C.,
  K\"{u}ttel, M., Mosley-Thompson, E., Overpeck, J.~T., Riedwyl, N., Schulz,
  M., Tudhope, A.~W., Villalba, R., Wanner, H., Wolff, E., and Xoplaki, E.
  (2009), \enquote{High-resolution palaeoclimatology of the last millennium: a
  review of current status and future prospects,} \textit{The Holocene}, 19,
  3--49.

\bibitem[{Jones and Mann(2004)}]{Jones2004}
Jones, P.~D. and Mann, M.~E. (2004), \enquote{Climate over past millennia,}
  \textit{Reviews of Geophysics}, 42, RG2002.

\bibitem[{K{\"o}rner(2008)}]{Korner2008}
K{\"o}rner, C. (2008), \enquote{Winter crop growth at low temperature may hold
  the answer for alpine treeline formation,} \textit{Plant Ecology \&
  Diversity}, 1, 3--11.

\bibitem[{K{\"o}rner and Paulsen(2004)}]{Korner2004}
K{\"o}rner, C. and Paulsen, J. (2004), \enquote{A world-wide study of high
  altitude treeline temperatures,} \textit{Journal of Biogeography}, 31,
  713--732.

\bibitem[{Li et~al.(2010)Li, Nychka, and Ammann}]{Li2010}
Li, B., Nychka, D.~W., and Ammann, C.~M. (2010), \enquote{The value of
  multiproxy reconstruction of past climate,} \textit{Journal of the American
  Statistical Association}, 105, 883--895.

\bibitem[{Masson-Delmotte et~al.(2013)Masson-Delmotte, Schulz, Abe-Ouchi, Beer,
  Ganopolski, Gonzalez-Rouco, Jansen, Lambeck, Luterbacher, Naish, Osborn,
  Otto-Bliesner, Quinn, Ramesh, Rojas, Shao, and
  Timmermann}]{Masson-Delmotte2013}
Masson-Delmotte, V., Schulz, M., Abe-Ouchi, A., Beer, J., Ganopolski, A.,
  Gonzalez-Rouco, J.~F., Jansen, E., Lambeck, K., Luterbacher, J., Naish, T.,
  Osborn, T., Otto-Bliesner, B., Quinn, T., Ramesh, R., Rojas, M., Shao, X.,
  and Timmermann, A. (2013), \enquote{Information from Paleoclimate Archives,}
  in \textit{Climate Change 2013: The Physical Science Basis. Contribution of
  Working Group I to the Fifth Assessment Report of the Intergovernmental Panel
  on Climate Change}, eds. Stocker, T.~F., Qin, D., Plattner, G.-K., Tignor,
  M., Allen, S.~K., Boschung, J., Nauels, A., Xia, Y., Bex, V., and Midgley,
  P.~M., Cambridge University Press, Cambridge, United Kingdom and New York,
  NY, USA.

\bibitem[{McShane and Wyner(2011)}]{McShane2011}
McShane, B.~B. and Wyner, A.~J. (2011), \enquote{A statistical analysis of
  multiple temperature proxies: are reconstructions of surface temperatures
  over the last 1000 years reliable?} \textit{The Annals of Applied
  Statistics}, 5, 5--44.

\bibitem[{Melvin and Briffa(2008)}]{Melvin2008}
Melvin, T.~M. and Briffa, K.~R. (2008), \enquote{A `signal-free' approach to
  dendroclimatic standardisation,} \textit{Dendrochronologia}, 26, 71--86.

\bibitem[{Melvin et~al.(2013)Melvin, Grudd, and Briffa}]{Melvin2013}
Melvin, T.~M., Grudd, H., and Briffa, K.~R. (2013), \enquote{Potential bias in
  `updating' tree-ring chronologies using regional curve standardisation:
  Re-processing 1500 years of {T}ornetr{\"a}sk density and ring-width data,}
  \textit{The Holocene}, 23, 364 -- 373.

\bibitem[{Osborne(1991)}]{Osborne1991}
Osborne, C. (1991), \enquote{Statistical calibration: A review,}
  \textit{International Statistical Review}, 59, 309--336.

\bibitem[{{PAGES 2k Consortium}(2013)}]{PAGES2kConsortium2013}
{PAGES 2k Consortium} (2013), \enquote{Continental-scale temperature
  variability during the past two millennia,} \textit{Nature Geoscience}, 6,
  339--346.

\bibitem[{Plummer(2003)}]{Plummer2003}
Plummer, M. (2003), \enquote{{JAGS}: A program for analysis of {B}ayesian
  graphical models using {G}ibbs sampling,} in \textit{Proceedings of the 3rd
  International Workshop on Distributed Statistical Computing, Vienna,
  Austria}.

\bibitem[{Schneider(2001)}]{Schneider2001}
Schneider, T. (2001), \enquote{Analysis of incomplete climate data: Estimation
  of mean values and covariance matrices and imputation of missing values,}
  \textit{Journal of Climate}, 14, 853--887.

\bibitem[{Tingley and Huybers(2010{\natexlab{a}})}]{Tingley2010}
Tingley, M. and Huybers, P. (2010{\natexlab{a}}), \enquote{A {B}ayesian
  algorithm for reconstructing climate anomalies in space and time. {Part 1}:
  Development and applications to paleoclimate reconstruction problems.}
  \textit{Journal of Climate}, 23, 2759--2781.

\bibitem[{Tingley et~al.(2012)Tingley, Craigmile, Haran, Li, Mannshardt, and
  Rajaratnam}]{Tingley2012}
Tingley, M.~P., Craigmile, P.~F., Haran, M., Li, B., Mannshardt, E., and
  Rajaratnam, B. (2012), \enquote{Piecing together the past: Statistical
  insights into paleoclimatic reconstructions,} \textit{Quaternary Science
  Reviews}, 35, 1--22.

\bibitem[{Tingley and Huybers(2010{\natexlab{b}})}]{Tingley2010a}
Tingley, M.~P. and Huybers, P. (2010{\natexlab{b}}), \enquote{A {B}ayesian
  algorithm for reconstructing climate anomalies in space and time. {Part II}:
  Comparison with the regularized expectation-maximization algorithm,}
  \textit{Journal of Climate}, 23, 2782--2800.

\bibitem[{Tolwinski-Ward et~al.(2011)Tolwinski-Ward, Evans, Hughes, and
  Anchukaitis}]{Tolwinski-Ward2011}
Tolwinski-Ward, S.~E., Evans, M.~N., Hughes, M.~K., and Anchukaitis, K.~J.
  (2011), \enquote{An efficient forward model of the climate controls on
  interannual variation in tree-ring width,} \textit{Climate Dynamics}, 36,
  2419--2439.

\bibitem[{Trenberth et~al.(2007)Trenberth, Jones, Ambenje, Bojariu, Easterling,
  Klein~Tank, Parker, Rahimzadeh, Renwick, Rusticucci, Soden, and
  Zhai}]{Trenberth2007}
Trenberth, K.~E., Jones, P.~D., Ambenje, P., Bojariu, R., Easterling, D.,
  Klein~Tank, A., Parker, D., Rahimzadeh, F., Renwick, J.~A., Rusticucci, M.,
  Soden, B., and Zhai, P. (2007), \enquote{Observations: surface and
  atmospheric climate change,} in \textit{Climate Change 2007: The Physical
  Science Basis. Contribution of Working Group I to the Fourth Assessment
  Report of the Intergovernmental Panel on Climate Change}, eds. Solomon, S.,
  Qin, D., Manning, M., Chen, Z., Marquis, M., Averyt, K.~B., Tignor, M., and
  Miller, H.~L., Cambridge University Press.

\bibitem[{Wahl and Smerdon(2012)}]{Wahl2012}
Wahl, E.~R. and Smerdon, J.~E. (2012), \enquote{Comparative performance of
  paleoclimate field and index reconstructions derived from climate proxies and
  noise-only predictors,} \textit{Geophysical Research Letters}, 39, L06703.

\bibitem[{Wigley et~al.(1987)Wigley, Jones, and Briffa}]{Wigley1987}
Wigley, T. M.~L., Jones, P.~D., and Briffa, K.~R. (1987), \enquote{Cross-dating
  methods in dendrochronology,} \textit{Journal of Archaeological Science}, 14,
  51--64.

\end{thebibliography}
